\documentstyle[preprint,aps,epsfig,epsf,floats]{revtex}
\begin{document}
\draft
\tighten

\newcommand{\st}{\scriptstyle}
\newcommand{\sst}{\scriptscriptstyle}
\newcommand{\mco}{\multicolumn}
\newcommand{\epp}{\epsilon^{\prime}}
\newcommand{\vep}{\varepsilon}
\newcommand{\ra}{\rightarrow}
\newcommand{\ppg}{\pi^+\pi^-\gamma}
\newcommand{\vp}{{\bf p}}
\newcommand{\ko}{K^0}
\newcommand{\kb}{\bar{K^0}}
\newcommand{\al}{\alpha}
\newcommand{\ab}{\bar{\alpha}}
\def\be{\begin{equation}}
\def\ee{\end{equation}}
\def\bea{\begin{eqnarray}}
\def\eea{\end{eqnarray}}
\def\CPbar{\hbox{{\rm CP}\hskip-1.80em{/}}}
\def\ap#1#2#3   {{\em Ann. Phys. (NY)} {\bf#1} (#2) #3.}
\def\apj#1#2#3  {{\em Astrophys. J.} {\bf#1} (#2) #3.}
\def\apjl#1#2#3 {{\em Astrophys. J. Lett.} {\bf#1} (#2) #3.}
\def\app#1#2#3  {{\em Acta. Phys. Pol.} {\bf#1} (#2) #3.}
\def\ar#1#2#3   {{\em Ann. Rev. Nucl. Part. Sci.} {\bf#1} (#2) #3.}
\def\cpc#1#2#3  {{\em Computer Phys. Comm.} {\bf#1} (#2) #3.}
\def\err#1#2#3  {{\it Erratum} {\bf#1} (#2) #3.}
\def\ib#1#2#3   {{\it ibid.} {\bf#1} (#2) #3.}
\def\jmp#1#2#3  {{\em J. Math. Phys.} {\bf#1} (#2) #3.}
\def\ijmp#1#2#3 {{\em Int. J. Mod. Phys.} {\bf#1} (#2) #3.}
\def\jetp#1#2#3 {{\em JETP Lett.} {\bf#1} (#2) #3.}
\def\jpg#1#2#3  {{\em J. Phys. G.} {\bf#1} (#2) #3.}
\def\mpl#1#2#3  {{\em Mod. Phys. Lett.} {\bf#1} (#2) #3.}
\def\nat#1#2#3  {{\em Nature (London)} {\bf#1} (#2) #3.}
\def\nc#1#2#3   {{\em Nuovo Cim.} {\bf#1} (#2) #3.}
\def\nim#1#2#3  {{\em Nucl. Instr. Meth.} {\bf#1} (#2) #3.}
\def\np#1#2#3   {{\em Nucl. Phys.} {\bf#1} (#2) #3.}
\def\pcps#1#2#3 {{\em Proc. Cam. Phil. Soc.} {\bf#1} (#2) #3.}
\def\pl#1#2#3   {{\em Phys. Lett.} {\bf#1} (#2) #3.}
\def\prep#1#2#3 {{\em Phys. Rep.} {\bf#1} (#2) #3.}
\def\prev#1#2#3 {{\em Phys. Rev.} {\bf#1} (#2) #3.}
\def\prl#1#2#3  {{\em Phys. Rev. Lett.} {\bf#1} (#2) #3.}
\def\prs#1#2#3  {{\em Proc. Roy. Soc.} {\bf#1} (#2) #3.}
\def\ptp#1#2#3  {{\em Prog. Th. Phys.} {\bf#1} (#2) #3.}
\def\ps#1#2#3   {{\em Physica Scripta} {\bf#1} (#2) #3.}
\def\rmp#1#2#3  {{\em Rev. Mod. Phys.} {\bf#1} (#2) #3.}
\def\rpp#1#2#3  {{\em Rep. Prog. Phys.} {\bf#1} (#2) #3.}
\def\sjnp#1#2#3 {{\em Sov. J. Nucl. Phys.} {\bf#1} (#2) #3.}
\def\spj#1#2#3  {{\em Sov. Phys. JEPT} {\bf#1} (#2) #3.}
\def\spu#1#2#3  {{\em Sov. Phys.-Usp.} {\bf#1} (#2) #3.}
\def\zp#1#2#3   {{\em Zeit. Phys.} {\bf#1} (#2) #3.}

\setcounter{secnumdepth}{2} 

\renewcommand{\floatpagefraction}{0.7}


\preprint{\parbox{5cm}{TTP95-39\ \\
HUTP-95/A041\ \\
hep-ph/9511268\\
November 1995 }}
\title{EXCLUSIVE SEMILEPTONIC TAU DECAYS\footnotemark[1]}

\author{J.H. K\"UHN AND E. MIRKES }

\address{Institut f\"ur Theoretische Teilchenphysik,
              Universit\"at Karlsruhe, D-76128 Karlsruhe, Germany}

\author{M. FINKEMEIER }

\address{Lyman Laboratory of Physics, Harvard University,
Cambrigde, MA 02138}

\maketitle

\begin{abstract}{
We review current issues in exclusive semileptonic tau decays.
We present the formalism of structure functions,
and then discuss predictions for final states with kaons,
for decays into four pions and for
radiative corrections to the decay into a
single pion.
}\end{abstract}

\footnotetext[1]{Invited talk given at the Europhysics
Conference on High Energy Physics, Brussels 27/7-2/8 1995;
presented by J.H. K\"uhn}

\section{Introduction}
The experimental and theoretical investigation of $\tau$ lepton decays
has received continuous attention since the $\tau$ has been
discovered nearly twenty
years ago. Branching ratios and distributions of the dominant modes
are now determined with a precision of a few permill, posing a serious
challenge for a refined theoretical treatment.
The purely leptonic decay rates can be evaluated from first principles,
providing thus a convenient calibration. The decay rate
$\Gamma(\tau\rightarrow\nu\pi)$ can be predicted using the measured
decay rate of the pion as input. Uncertainties arise at the permille level and
are induced through the model dependence of radiative corrections. CVC
allows to relate vector current decays e.g. into two or four pions
to the corresponding cross sections measured in electron positron collisions.
However, already now this approach is limited by the error of the input data.
In the case of the two pion mode, the $\tau$ decay rate with a relative error
below one \% is already now about
a factor  five  more precise than the predictions based on
$\sigma(e^+e^-\rightarrow\pi\pi)$. For the remaining modes, induced by the
axial current or by Cabibbo suppressed channels, a variety of theoretical
assumptions must be employed, based on symmetry relations between different
amplitudes which can be derived from isospin or $SU(3)$, or on dynamical
constraints deduced from chiral Lagrangians at small momentum transfer.
To saturate the form factors in the region of large $Q^2$ by resonances with
momentum independent coupling is a natural choice in the
context of the vector dominance model (VDM)
but perhaps the most problematic assumption,
which has to be tested by measurements of differential distributions.
A particular powerful tool is provided by the analyses of
angular distributions. The relevant information is conveniently encoded in
structure functions\cite{km}
which, in turn, allow to reconstruct the form factors.
$\tau$ decays are therefore a unique tool to study hadron physics in the
low momentum region in order to test a variety of theoretical approaches and
to derive resonance parameters which are not accessible elsewhere.
An improved description of the hadronic matrix elements allows,
furthermore, to exploit all (or at least most) channels for an analysis
of the $\tau$ polarization $P_{\tau}$  in $Z$ decays. Decays into a pion have
been employed for this purpose from the very beginning; as a consequence
of the improved understanding of the relevant distributions two and three pion
modes are now equally important for the analysis.
Since all multi-pion modes have equal (and maximal) analyzing power once the
decay matrix element is known\cite{k} this is clearly a worthwhile aim.
Another reason to improve the  theoretical description of $\tau$
production and decay is the
search for new physics. Subtle deviations from the Standard Model like
the influence
of a charged Higgs boson
on the question of universality  could be amplified by the masses
of the fermion involved in the reaction\cite{higgs}.
Last not least: only a steady improvement of the theoretical understanding
allows to develop the Monte Carlo program TAUOLA\cite{tauola},
which is used to simulate $\tau$ decays in

practically all $e^+e^-$ experiments
at LEP, SLC and CESR.

In this contribution a variety of topics and recent developments will
be treated. In section 2 the formalism for the relation between form factors
and structure functions and the transition to angular distributions will be
reviewed\cite{km} and the basic assumptions inherent in the theoretical
models used to predict multimeson amplitudes will be presented\cite{ks}.
In section 3 predictions for final states involving one or several
kaons will be discussed, with emphasis on a recent update\cite{kaonnew}
of the theoretical predictions. Four pion modes will be
discussed in section
4, which is  based on ref.~[7].
Radiative corrections\cite{df1} to
$\tau\rightarrow\nu_{\tau}\pi(\gamma)$ will be discussed in the last,
fifth section of this review. \nopagebreak{
\section{Form Factors, Structure Functions \protect\newline and
$\Gamma(\tau\rightarrow \nu_{\tau}3\pi)$}
The matrix element for the $\tau$ decay into a multimeson final state
can be written in the following form}
\begin{equation}
{\cal M}(\tau\rightarrow\nu\,\,\mbox{had})=
\frac{G_F}{\sqrt{2}}\bar{u}(l^\prime)\gamma_{\mu}(1-\gamma_5)u(l)\,J^{\mu}
\end{equation}
The hadronic current $J^{\mu}$ is built up from the momenta of the mesons.
For the three meson state it is characterized by four
independent form factors.
A convenient choice reads as follows\cite{km}
\begin{eqnarray}
J^{\mu}&=&\left(g^{\mu\nu}-\frac{Q^{\mu}Q^{\nu}}{Q^2}\right)
          \left[ \,(q_1-q_3)_{\nu}\,F_1\,+\,(q_2-q_3)_{\nu}\,F_2\,\right]
        \nonumber\\
   & &+\,\,i\,
  \epsilon^{\mu\alpha\beta\gamma}q_{1\,\alpha}q_{2\,\beta}q_{3\,\gamma}\,F_3
  + Q^\mu\,F_4
\end{eqnarray}
$F_1$ and $F_2$ determine the spin one component of the axial current induced
amplitude, $F_4$ the spin zero part which is obviously given by the
matrix element of the divergence of the axial current. The vector current
induced amplitude is responsible for the formfactor $F_3$.
All form factors may contribute in the general three meson case.
The three pion case, however, allows for significant simplifications.
$G$-parity implies $F_3=0$, Bose symmetry relates $F_1$ and $F_2$ and
PCAC leads to $F_4=0$. Instead of 16 independent real functions
which characterize the general hadronic tensor
$H^{\mu\nu}=J^{\mu}J^{\ast \nu}$ one thus deals with four functions of
$Q^2,s_1,s_2$ only. A convenient basis has been found in ref.~[1]
which is related to the form factors in a straightforward way.
In the three pion rest frame, and with the $x,y$ coordinates alligned with the
$\vec{q}_1,\vec{q}_2$ plane (see Fig.~1), these four ``structure functions''
are given by
\begin{eqnarray}
W_{A}  &=&   \hspace{3mm}(x_{1}^{2}+x_{3}^{2})\,|F_{1}|^{2}
           +(x_{2}^{2}+x_{3}^{2})\,|F_{2}|^{2} \nonumber\\
       && \hspace{3mm}    +2(x_{1}x_{2}-x_{3}^{2})\,
           \mbox{Re}\left(F_{1}F^{\ast}_{2}\right)
                                   \nonumber \\
W_{C}  &=&  \hspace{3mm} (x_{1}^{2}-x_{3}^{2})\,|F_{1}|^{2}
           +(x_{2}^{2}-x_{3}^{2})\,|F_{2}|^{2}\nonumber\\
       &&\hspace{3mm}    +2(x_{1}x_{2}+x_{3}^{2})\,
           \mbox{Re}\left(F_{1}F^{\ast}_{2}\right)
                                \label{walldef}   \\
W_{D}  &=&  \hspace{3mm}2\left[ x_{1}x_{3}\,|F_{1}|^{2}
           -x_{2}x_{3}\,|F_{2}|^{2}\right. \nonumber\\
       &&\hspace{3mm}      \left.     +x_{3}(x_{2}-x_{1})\,
           \mbox{Re}\left(F_{1}F^{\ast}_{2}\right)\right]
                                   \nonumber \\
W_{E}  &=& -2x_{3}(x_{1}+x_{2})\,\mbox{Im}\left(F_{1}
                    F^{\ast}_{2} \right) \nonumber
\end{eqnarray}
The variables $x_i$ are defined by
$
x_{1}= q_{1}^{x}-q_{3}^{x},\,
x_{2}= q_{2}^{x}-q_{3}^{x},\,
x_{3}= q_{1}^{y}=-q_{2}^{y},\,
$
where $q_i^{x}$ ($q_i^{y}$) denotes the $x$ ($y$) component of the momentum of
meson $i$ in the hadronic rest frame.
$W_A$ governs the rate and the distributions in the Dalitz plot,
the remaining functions determine the angular distribution.
If the $\tau$ is unpolarized and its rest frame is reconstructed, the angular
distribution of the three pions with respect to the $\tau$ momentum
(as seen from the 3 pion rest frame) is given by
\begin{eqnarray}
\frac{dN}{d\cos\beta d\gamma}&\sim&
\left[ \left(1-{m_\tau^2}/{Q^2}\right)(1-\cos^2\beta)
      +2{m_\tau^2}/{Q^2}\right]\,W_A\nonumber\\
&&\hspace{-2cm}
     -\left(1-{m_\tau^2}/{Q^2}\right)\sin^2\beta\,\,
     (\cos 2\gamma\,W_C-\sin 2\gamma\, W_D)\\
&& + 2\,\,\cos\beta\,W_E\nonumber
\end{eqnarray}
with the definition of the angles indicated in Fig.~1.
In ref.~[1] it has been demonstrated that these structure
functions can also be measured in present experiments
despite the fact that
the neutrino momentum is missing and hence the analysis has to be
performed without knowledge of the $\tau$ rest frame.
Predictions for the structure functions have been made
in ref.~[1] (see also Fig.~2) based on the model of ref.~[5], and in fact
first comparisons between theory and experiment have been performed
in ref.~[9]
\setlength{\unitlength}{0.7mm}
\begin{figure}[hbt]
\vspace{5mm}
\begin{picture}(150,165)(-50,1)
\mbox{\epsfxsize12.0cm\epsffile{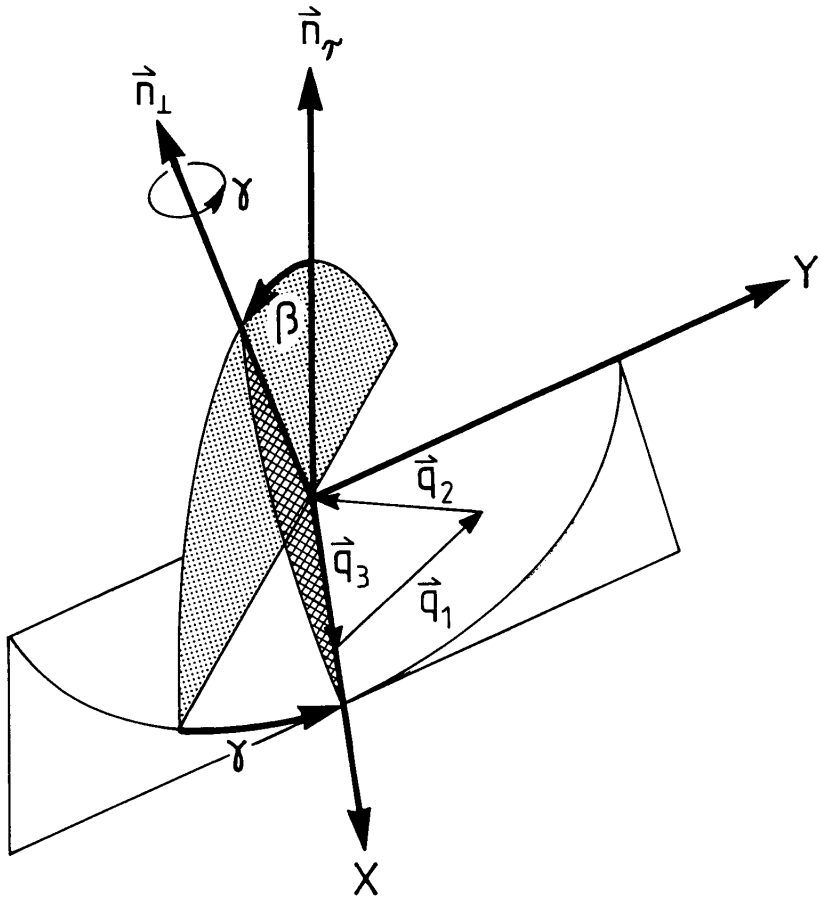}}
\end{picture}
\caption{Definition of the angles $\beta$ and $\gamma$.}
\end{figure}
\setlength{\unitlength}{0.7mm}
\begin{figure}[hbt]               \vspace*{-2cm}
\begin{picture}(150,165)(-30,1)
\mbox{\epsfxsize12.0cm\epsffile[30 80 430 500]{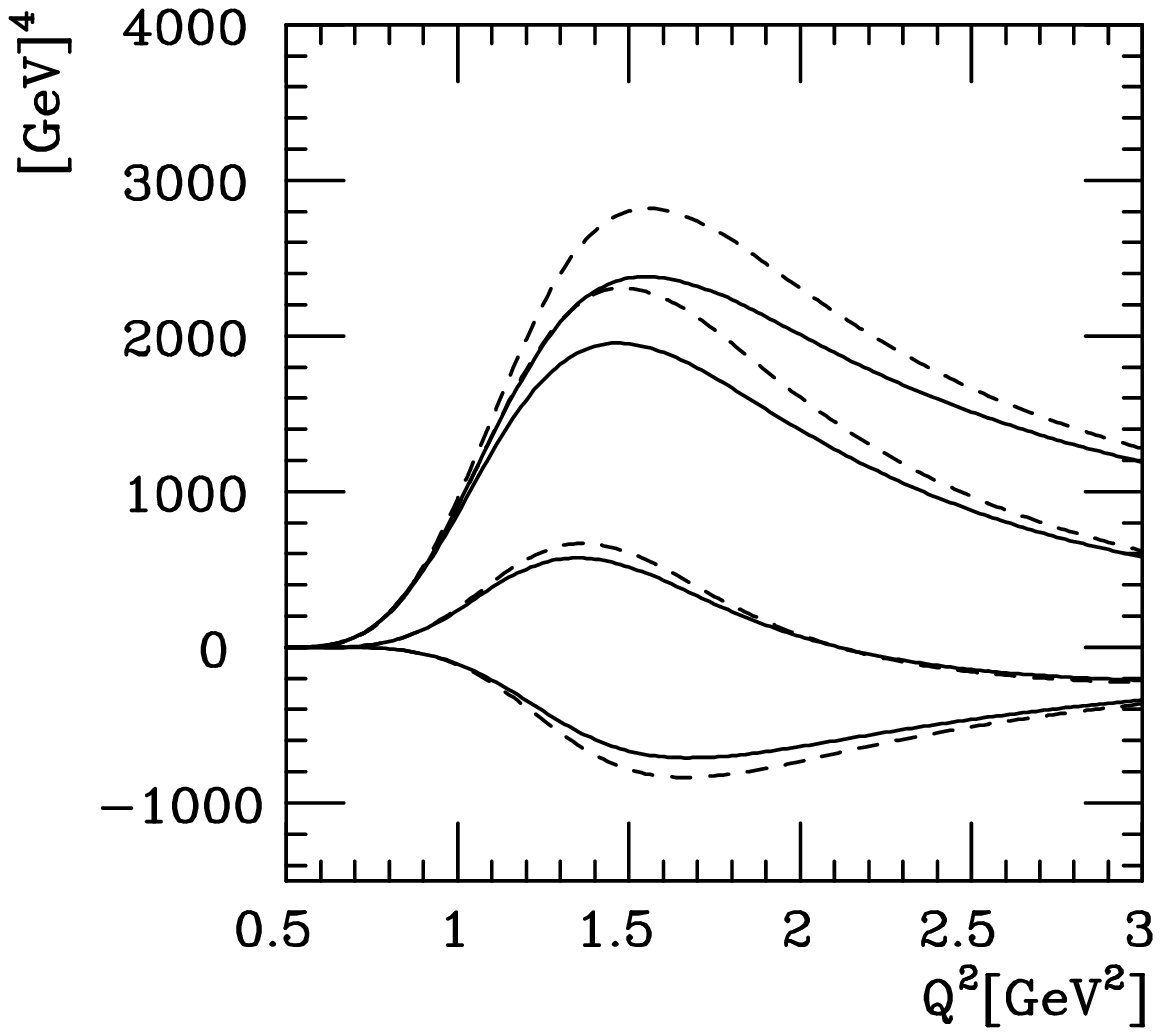}}
\end{picture}
\caption{Spin one hadronic structure functions
         $W_A, W_C, W_D$ and $W_E$ (from top to bottom) for
         $\tau\rightarrow \nu3\pi$ as a function of $Q^2$.
         Results are shown for two sets of $a_1$ parameters:
         $m_{A_1}=1.251$ GeV, $\Gamma_{a_1}= 0.599$ GeV (solid)
         and
         $m_{A_1}=1.251$ GeV, $\Gamma_{a_1}= 0.550$ GeV (dashed) }
\end{figure}
In this model (earlier variants based on a similar approach
can be found in ref.~[10])

the normalization of the
form factor is determined in the chiral limit:
\begin{equation}
F_1=F_2=\frac{2\sqrt{2}}{3f_{\pi}}\cos\theta_c
\end{equation}
For large $Q^2,s_1$ and $s_2$ these formfactors are modulated by
Breit Wigner resonances in the $3\pi$ and $2\pi$ channel:
\begin{eqnarray}
F_1(Q^2,s_2)&=& \frac{2\sqrt{2}}{3f_{\pi}}\cos\theta_c
BW_{A_1}(Q^2)\,BW_{\rho}(s_2)\\
F_2(Q^2,s_1)&=& \frac{2\sqrt{2}}{3f_{\pi}}\cos\theta_c
BW_{A_1}(Q^2)\,BW_{\rho}(s_1)
\end{eqnarray}
More detailled studies, testing the magnitude of amplitudes
induced by $F_3$ and $F_4$, are possible. Since these affect the
angular distributions through interference terms, they should
be well accessible in measurements, already with the
statistics of ongoing experiments.

\section{Final States with Kaons [6,11]}
A rich variety of final states has become accessible
recently with improved particle identification: modes with
one or two kaons have been analyzed and even the three kaon
final state is kinematically accessible in $\tau$ decays.
Considering the multitude of charge assignments, even the
bookkeeping for the various reactions becomes tedious.
The ansatz for the $K\pi$ channel is still straightforward.
Dominated by $K^\ast$ and its radial excitation ${{K^\ast}'}$,
the form factor is constructed in close similarity to the one
of the two pion mode, which proceeds through the $\rho$ and $\rho^\prime$
mesons, of course modified by proper $SU(3)$ factors.
In fact, the relative coupling strength of ${{K^\ast}}$ and ${K^\ast}'$
is identical to the one of $\rho$ and $\rho^\prime$ in the two pion case.

The choice of a specific form for the $K\pi$ resonant amplitude
is an important ingredient for the $K\pi\pi$ and $KK\pi$

channels\cite{kaonnew}
to be discussed now.
The construction of the relevant form factors proceeds in close
similarity to the approach discussed in section~3 for the three pion
channel. For small momenta the form factors
 $F_1$ and $F_2$ are again derived
from chiral Lagrangians.
In addition, one finds in the same limit a non-vanishing $F_3$.
For example,
\begin{equation}
F_3^{(K^-\pi^-K^+)}(q_i=0)=\frac{-1}{2\sqrt{2}\pi^2f_{\pi}^3}
\end{equation}
and similarly for the other kaonic modes
(except for $K^-\pi^0K^0$ and $\pi^0\pi^0K^-$ where $F_3$
vanishes in the chiral limit).
This term is induced by the celebrated Wess-Zumino anomaly and has
originally been discussed in the context of $\tau$ decays in ref.~[12]
Vector as well as axial current induced amplitudes are present leading to
five additional  non-vanishing structure functions
(compared to the three pion mode)
and to more complicated angular distributions\cite{km}.
The incorporation of resonances in the three meson channel and in various
two meson channels proceeds essentially as in the three pion mode.
A number of additional complications do however arise:
The form factor $F_2\, (F_3)$ vanishes in the soft meson limit
for $K^-\pi^0K^0\, (K^-\pi^0K^0,\,\pi^0\pi^0K^-)$.
On the other hand it is evident that spin one resonances are present
in all three subchannels
$K^-K^0 \equiv \rho$, $K^-\pi^0\equiv K^{- \star}$ and
$K^0\pi^0\equiv K^{0\star}$.
Hence $F_2\,(F_3)$ must be different from zero for non-vanishing
$q_i$.
The resulting choice of amplitudes is treated in ref.~[6]
These relations provide important cross checks on the
consistency of different measurements.

Depending on the details of
the experimental setup, one may either observe $K^0$ and $\bar{K}^0$
(if the kaons interact in the detector) or
$K_S$ and $K_L$, if only their decay products are observed.
The rates for $K_L\pi^-K_L$ and $K_S\pi^-K_S$ are identical,
the rate for $K_S \pi^- K_L$ is about a factor 2.1 higher.
Note that the first relation is a strict consequence of
CP symmetry, the second one depends on the dynamics of the decay
amplitude. Table 1 displays the theoretical predictions for the
vector and axial vector induced rate separately and compares them with
the experimental results presented at this conference\cite{exp}
[using  $B(\tau\rightarrow\nu\nu_ee)= 17.8\%$].
Experiment and theory are in satisfactory agreement.


\begin{table}
\caption{Predictions for the normalized decay widths
$\Gamma(abc)/\Gamma_e$ and the branching
ratios ${\cal B}(abc)$ for the various channels.
The contribution from the vector current is listed in column 3 and
available experimental data are listed in column 5.
}
\label{tab3}
$$
\begin{array}{c@{\quad}c@{\quad}c@{\quad}c@{\quad}c}
\hline \hline
\begin{array}{c}
\mbox{channel}
\\
(abc)
\end{array}
& \! \! \! \!
\left( \frac{\Gamma{(abc)}}{\Gamma_e} \right) & \! \! \! \!
\left( \frac{\Gamma{(abc)}}{\Gamma_e} \right)_{V} & \! \! \! \!
\begin{array}{c}
{\cal B}{(abc)}
\\ {} [\%] \end{array}
& \! \! \! \!
\begin{array}{c}
{\cal B}{(abc)}^{(expt.)}
\\ {} [\%] \end{array}
 \\
\hline
\\[-4mm]

\\K^- \pi^- K^+          & .011   & .0045  & .20  & ( .20 \pm .07)  \\
K^0 \pi^- \overline{K^0} & .011   & .0045  & .20   &                 \\
K_S \pi^- K_S            & .0027  & .0008  & .048  &  (.021\pm .006)  \\
K_S \pi^- K_L            & .0058  & .0029  & .10   &                 \\
K^- \pi^0 K^0            & .0090  & .0032  & .16   &  (.12 \pm .04)  \\
\eta \pi^- \pi^0         & .0108  & .0108  & .19   &  (.170 \pm .028)  \\[1mm]
\hline \\[-4mm]
\pi^0 \pi^0 K^-          & .0080  & .0007  & .14   &  (.09 \pm .03)   \\
K^-   \pi^-\pi^+         & .043   & .0043  & .77   &  (.40 \pm .09)   \\
\pi^-\overline{K^0}\pi^0 & .054   & .0058  & .96   &  (.41 \pm .07)   \\[1mm]
\hline \hline
\end{array}
$$
\end{table}

\section{$\tau$ Decays into Four Pions, CVC and
$\lowercase{e}^+\lowercase{e}^-$ Annihilation.}

In principle the total and differential decay rate into four
pions can be related to measurements of the cross section for
$e^+e^-$ annihilation into four pions.
\begin{eqnarray}
d\Gamma(\tau\rightarrow\nu\pi^-3\pi^0) &\sim&
      d\sigma(e^+e^-\rightarrow 2\pi^-2\pi^+)\nonumber\\
d\Gamma(\tau\rightarrow\nu2\pi^-\pi^+\pi^0) &\sim&
      d\sigma(e^+e^-\rightarrow 2\pi^-2\pi^+) \\
        &+&
      2\,d\sigma(e^+e^-\rightarrow \pi^+\pi^-2\pi^0)\nonumber
\end{eqnarray}
However, already now the $\tau$ decays rate is comparable in
precision to the $e^+e^-$ data. In particular the data for
$e^+e^-\rightarrow \pi^+\pi^-2\pi^0$ exhibit significant discrepancies
between different experiments (see Fig.~3)
which are not apparent from the theoretical predictions based on
simply averaging the different sets of data\cite{eidelmann}.
The model developed in ref.~[7] prefers the experimental
results with the smaller cross section. Precise data from
$\tau$ decays can help to resolve the problem.
The model for the four pion channel was originally formulated
in refs.~[15,4] and has been recently improved and expanded
in ref.~[7].
The amplitude is again normalized to the predictions in the chiral limit,
supplemented by appropriate resonances in the two pion $(=\rho)$ and
three pion $(=a_1)$ channel. In addition the anomalous
$\pi\omega(\rightarrow3\pi)$
mode is added, with a $\rho\omega\pi$ coupling strength derived

from experimental information on $\omega\rightarrow\pi\gamma$ and
VDM\cite{d}.
The $\omega\pi$ contribution to the $4\pi$ is expected to proceed via
a vector current. However, a violation of $G$-parity would allow the
$\omega\pi$ system to be in an axial vector state, which could be
revealed by an analyses of the angular distribution in the $\omega\pi$
mode as introduced in ref.~[17].

Predictions for the differential distributions of the model are in reasonable
agreement with the experimental results.
%
%
%
%
 \begin{figure}
 \begin{flushleft}
 \mbox{$\displaystyle  \sigma \, [\mbox{nb}]$}
 \end{flushleft}
 \begin{center}
 \mbox{\epsfig{figure=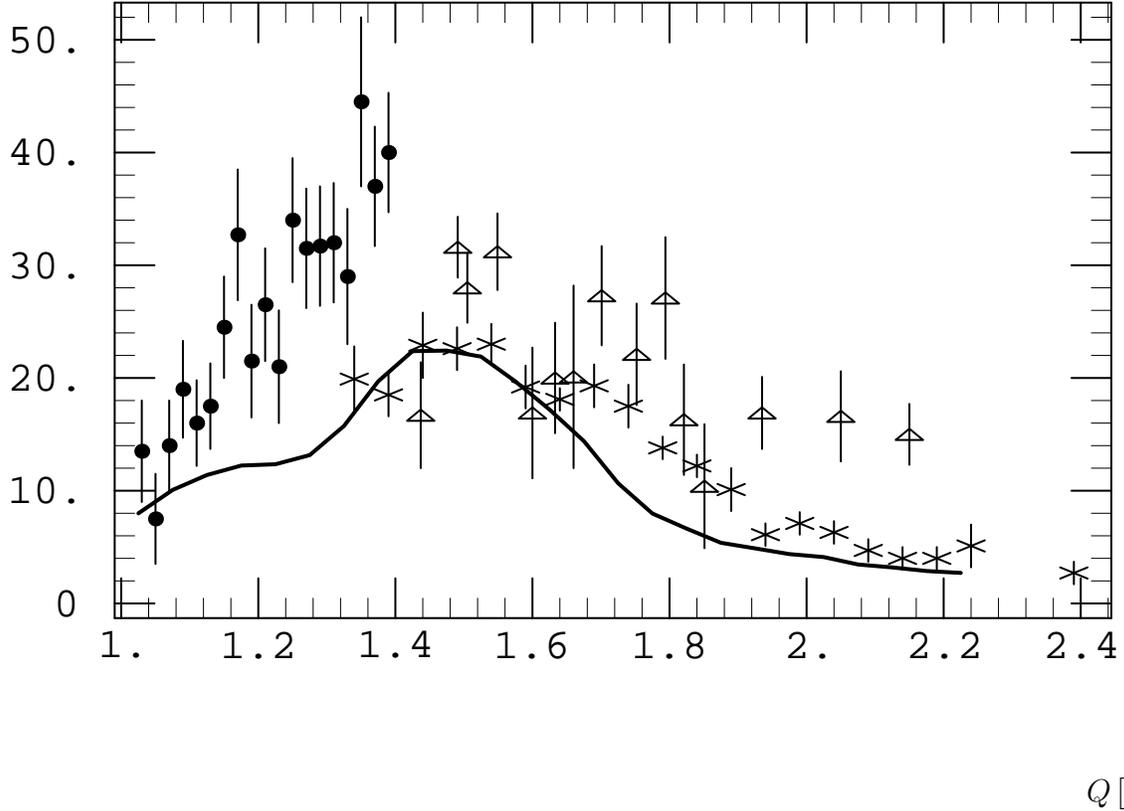,width=16cm,%
 bbllx=2.cm,bblly=8.cm,bburx=20.cm,bbury=20.cm}}
 \end{center}
 \begin{flushright}
 \mbox{$\displaystyle Q \, [\mbox{GeV}] $}
 \end{flushright}
\caption{
Cross section for $e^+ e^- \to \pi^+ \pi^- 2 \pi^0$: Data
from Nowosibirsk (dots with error bars) \protect \cite{nowo},
from Frascati (triangles) \protect \cite{frasci} and
from Orsay (crosses) \protect \cite{orsay}, and the prediction
from the model in ref.~[7] (solid line)}
 \end{figure}

\section{Radiative Corrections to $\tau\rightarrow\nu\pi(\gamma)$}
Neglecting radiative corrections, the decay rate of the $\tau$ lepton into
a pion $\Gamma(\tau\rightarrow\pi\nu)$
can be unambiguously predicted from the pion decay rate
 since both quantities are proportional to the pion decay
constant. However, virtual as well as real photon emision are sensitive to
different regions of the respective form factors,
and hence introduce a finite correction  $\delta R_{\tau/\pi}$ in the ratio
\begin{eqnarray}
R_{\tau/\pi} & = &
\frac{\Gamma(\tau \to \nu_\tau \pi (\gamma))}
     {\Gamma(\pi \to \nu_\mu \mu (\gamma))}
\nonumber \\
& = &\frac{m_\tau^3}{2 m_\pi m_\mu^2}
   \frac{(1 - m_\pi^2/m_\tau^2)^2}{(1 - m_\mu^2/m_\pi^2)^2}
  \Big(1 + \delta R_{\tau/\pi} \Big)
\end{eqnarray}
A detailled study of these form factors and their impact on $R_{\tau/\pi}$
has been performed in ref.~[8].
The behaviour of the form factors at very small momentum transfers
is fixed by low energy theorems like the axial anomaly and by
experimental results obtained in radiative pion decay.
In the region of intermediate momenta phenomenological parameterizations
inspired by vector meson dominance are employed. The uncertainty in the final
prediction is dominated by the ignorance of the form factor in this region.
This ansatz is sufficient to predict the photon energy spectrum over
the full kinematical region. As demonstrated in Fig.~4,

the model prediction is in fair agreement with the prediction from

PHOTOS\cite{photos},
which is based on a semiclassical approximation.

%
\begin{figure}
\label{figu4}
\begin{center}
%
\fbox{   \epsfig{figure=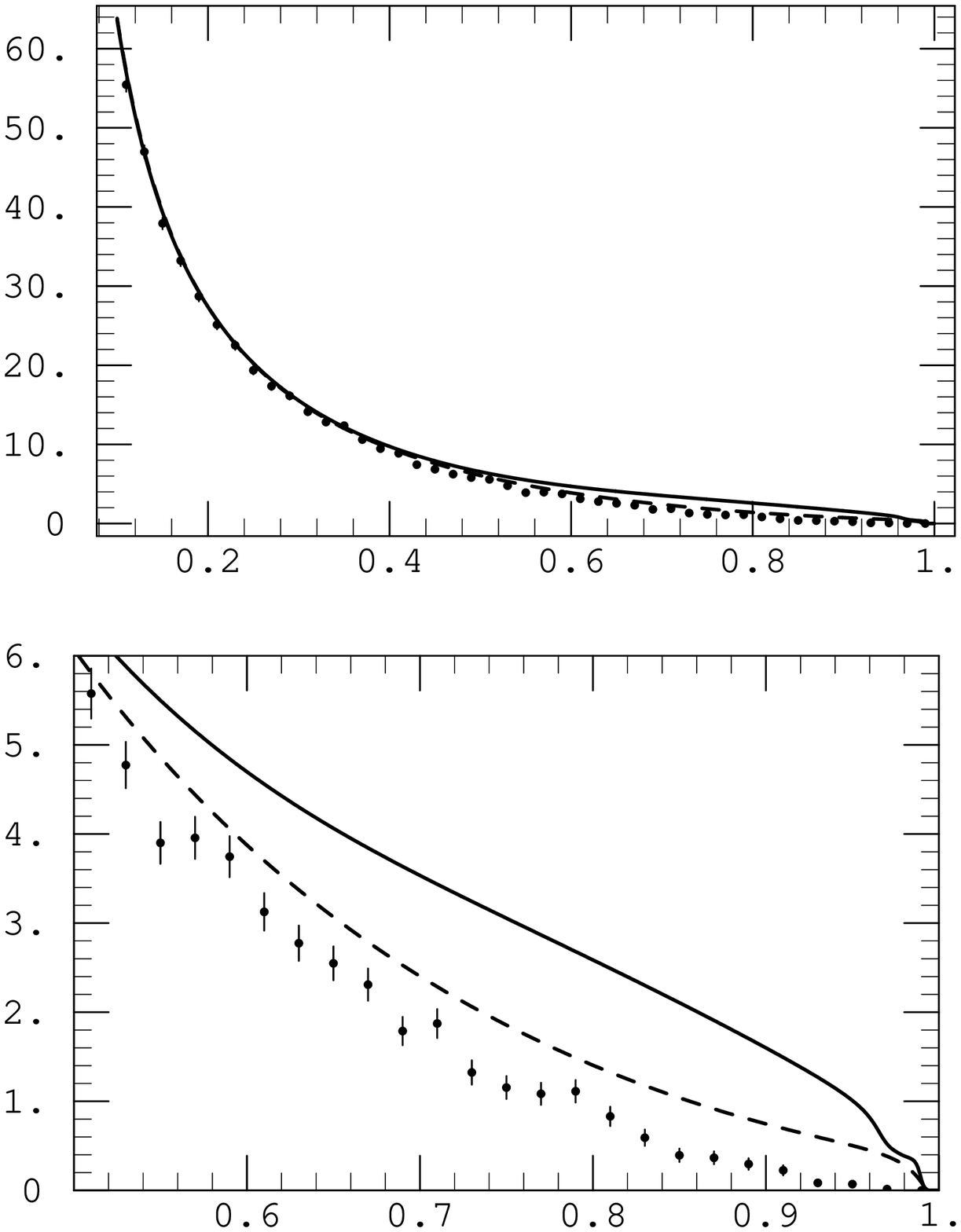,width=7cm,%
bbllx=2.5cm,bblly=3.0cm,bburx=19.4cm,bbury=25.5cm}}
\unitlength0.5cm
\begin{picture}(0,0)
   \put(-11.5,18.){\makebox(0,0)[l]{
   \scriptsize {\bf (a)} $0.1 \leq x \leq 1.0$ }}
   \put(-11.5,9.0){\makebox(0,0)[l]{\scriptsize{\bf (b)} Close-up of (a) %
    with $0.5 \leq x \leq 1$}}
  \put(-2.5,9.7){\makebox(0,0)[l]{  $ x $ }}
  \put(-6,16.){\makebox(0,0)[l]{ $ \frac{ 10^{3}}{
\Gamma_{\tau\to\pi\nu_\tau}}%
  \frac{d\Gamma}{dx}$}}
  \put(-2.5,0.6){\makebox(0,0)[l]{  $ x $ }}
  \put(-6,7.){\makebox(0,0)[l]{ $ \frac{ 10^{3}}{ \Gamma_{\tau\to\pi\nu_\tau}}%
  \frac{d\Gamma}{dx}$}}
\end{picture}
  \end{center}
\caption{Photon energy spectrum of the decay $\tau\to\pi\nu_\tau\gamma$:
Total prediction (solid), internal bremsstrahlung alone
(dashed) and the prediction from PHOTOS
(dots with statistical error bars)}
\end{figure}
%
The virtual radiation
extends over the full kinematical range with $Q^2$ even up to $M_Z^2$.
For large momentum transfer the quark content of the pion plays the dominant
role.  Integrating over the quark-antiquark wave function leads to a
short distance contribution proportional to $f_\pi$. Varying the
model parameters and the cutoff  $\mu^2_{cut}$ which separates long and
short distance contributions one arrives at a result for
$\delta R_{\tau/\pi}$ and equally important at an estimate of the theoretical
uncertainty of the prediction
\begin{equation}
\delta R_{\tau/\pi}= (0.16 \pm 0.14) \%
\end{equation}
Similarly one obtains a result for the kaon mode
\begin{equation}
\delta R_{\tau/K}= (0.90 \pm 0.22) \%
\end{equation}
The prediction for the pion mode should be contrasted with a recent estimate
by Marciano and Sirlin\cite{Mar93},
which in terms of $R_{\tau/\pi}$ reads
\begin{equation}
\delta R_{\tau/\pi}= (0.67 \pm 1.) \%
\end{equation}
where the error reflects the authors' estimate of the missing long
distance corrections which are not included in their work.
{}From the work of ref.~[8] it is evident that the $\tau$ decay rate
to $\nu\pi(\gamma)$ and $\nu K (\gamma)$ can provide a unique test of lepton
universality. Extracting the decay rate from ${\cal B}(\nu\pi(\gamma))$ and
the $\tau$ lifetime, the equality of $\tau$ and muon coupling to the
charged current can be tested with a precision up to one permille.

\section{Summary}
Exclusive semileptonic $\tau$ decays offer a unique tool to test
the Standard Model in the low energy region. Resonance parameters and
form factors can be determined with results complementary  to those
obtained in low energy $e^+e^-$ machines. The $\tau$ polarization
can be determined from all decay modes with optimal analyzing power.
Isospin and $SU(3)$ symmetry can be investigated. Lepton universality
can be tested with a precision of about one permille.
Deviations from the Standard Model induced by loop effects
or additional tree level contributions {\it e.g.} from a two Higgs
doublet model can thus be explored in an interesting range of parameters.

\section*{ACKNOWLEDGEMENTS }
%
The work of M.F. was supported

by the Deutsche Forschungsgemeinschaft (Grant Fi 635/1-1).

Further support was provided by the

National Science Foundation (Grant PHY-9218167).
%

\end{document}